\documentclass[10,pra,a4paper,aps,twocolumn,superscriptaddress]{revtex4-1}

\usepackage{amsmath,amsthm,amssymb,amsfonts}
\usepackage{float}
\usepackage{graphicx}
\usepackage{multirow}
\usepackage{url}
\usepackage[usenames,dvipsnames]{xcolor}
\usepackage[breaklinks=true]{hyperref}
\usepackage{bbold}
\usepackage{cleveref} 
\usepackage[normalem]{ulem}
\usepackage{comment}
\usepackage[normalem]{ulem}

\usepackage[utf8]{inputenc}
\usepackage[T1]{fontenc}
\usepackage{tikz}
\usepackage{tensor}

\theoremstyle{definition}

\newtheorem*{exmp*}{Example}

\newcommand{\ket}[1]{| #1 \rangle}
\newcommand{\bra}[1]{\langle #1 |}

\begin{document}
\title{Feynman path sum approach for simulation of linear optics}

\author{W. F. Balthazar}

\affiliation{Instituto Federal do Rio de Janeiro,  R. Antônio Barreiros, 212, Volta Redonda - RJ, 27213-100, Brasil}
\affiliation{International Iberian Nanotechnology Laboratory (INL), Av. Mestre José Veiga s/n, 4715-330 Braga, Portugal}

\affiliation{Programa de Pós-graduação em Física, Instituto de F\'isica, Universidade Federal Fluminense, Av. Gal. Milton Tavares de Souza s/n, Niter\'oi, RJ, 24210-340, Brazil}

\author{Q. M. B. Palmer}
\affiliation{Quantum Engineering Technology Labs, H.H. Wills Physics Laboratory and Department
of Electrical and Electronic Engineering, University of Bristol, Bristol, United Kingdom}
\author{A. E. Jones}
\affiliation{Quantum Engineering Technology Labs, H.H. Wills Physics Laboratory and Department
of Electrical and Electronic Engineering, University of Bristol, Bristol, United Kingdom}
\author{J. F. F. Bulmer}
\affiliation{Quantum Engineering Technology Labs, H.H. Wills Physics Laboratory and Department
of Electrical and Electronic Engineering, University of Bristol, Bristol, United Kingdom}
\author{E. F. Galv\~ao}

\affiliation{International Iberian Nanotechnology Laboratory (INL), Av. Mestre José Veiga s/n, 4715-330 Braga, Portugal}
\affiliation{Instituto de F\'isica, Universidade Federal Fluminense, Av. Gal. Milton Tavares de Souza s/n, Niter\'oi, RJ, 24210-340, Brazil}
\date{}
\begin{abstract}

The Feynman path integral formalism has inspired the development of memory-efficient and parallelizable classical algorithms for simulating quantum computers. We adapt this approach for the calculation of probability amplitudes of linear-optical boson sampling experiments, which involve Fock-state inputs, linear optical circuits, and photo-detection at the output. We describe this simulation method and compare it with alternative approaches. Additionally, we implement a Linear-Optical Feynman Path simulator in open-source C code, enhancing its performance using tensor contraction techniques. Our method is benchmarked for low-depth linear optical circuits, where it offers advantages in runtime and memory efficiency.

\end{abstract}

\maketitle

\section{Introduction}

There are two main approaches to the classical simulation of general quantum circuits. The first, sometimes called the Schr{\"o}dinger approach, involves storing and updating the quantum state after each applied gate. In this approach, both space (memory) and runtime increase exponentially with the number of qubits. The second approach, inspired by the Feynman path integral formulation of quantum dynamics~\cite{feynman48spacetime}, also incurs a runtime that is exponential in the number of qubits, but using only polynomially increasing space \cite{AaronsonC16}. Feynman path sums are easy to parallelize, allowing, for example, for the recent simulation~\cite{PanCZ22} of the 2019 Google Quantum AI quantum advantage experiments~\cite{Arute19}. An interpolation between the Schr{\"o}dinger and the Feynman approaches allows trade-offs between the advantages of each approach~\cite{AaronsonC16}. Tensor contraction techniques with different contraction sequences also allow for trade-offs between runtime and memory~\cite{MarkovS08}.

So far, quantum computational advantage has arguably been demonstrated using only two restricted models of quantum computation: random circuit sampling \cite{Arute19, Gao25}, and boson sampling \cite{zhong2020quantum, madsen2022quantum}. The boson sampling~\cite{Aaronson11} model, of interest to us here, features Fock-state interference in multi-mode linear interferometers, with particle-counting detectors at the output. Many experimental realizations have been reported -- for a review, see~\cite{Brod19review}. Recently, an atomic boson sampler has been used to implement this restricted model of quantum computation in the high complexity regime \cite{young2024atomic}. A variant that uses Gaussian state inputs~\cite{Hamilton17} has been demonstrated in impressive experiments that claim to outperform classical supercomputers by a large margin \cite{zhong2020quantum, liu25robust}.

In strong simulation, the probability amplitude of a candidate setup is calculated. In our case, we consider a linear-optical circuit whose amplitude is determined by the permanent of a matrix associated with the interferometer design and the choice of input and output states \cite{scheel2004permanents, Aaronson11}. These permanent calculations are crucial for validating experimental outcomes and benchmarking quantum devices. The most efficient general-purpose algorithm for this strong simulation is Ryser's algorithm with Gray code ordering \cite{Ryser1963}, whose runtime scales as $O(n2^n)$, where $n$ is the matrix size. However, some algorithms can offer advantages by exploiting the sparseness of matrices associated with a shallow circuit~\cite{CifuentesP16,novak2024laplace}.

Here, we propose a Feynman path sum approach for strong simulation, i.e., exact calculation of probability amplitudes for Fock-state boson sampling.
We discuss the method's performance for simulating the evolution of photons through layers of locally-connected beam splitters (BS). Furthermore, we highlight how tensor contraction techniques can improve the efficiency for simulating interferometers with specific connectivity structures~\cite{MarkovS08}, albeit at the cost of increased memory requirements. 

We benchmark our method by comparing it to Ryser's algorithm \cite{Ryser1963} using Gray code ordering \cite{gray1953pulse} and an implementation of the Cifuentes-Parrilo algorithm that is faster for matrices with small treewidth~\cite{CifuentesP16}, a quantity related to the sparsity of the unitary matrix representing the 
interferometer. We illustrate our approach with simulations of a commonly used planar 
interferometer design proposed by Clements \textit{et al.} \cite{Clements16}, looking in particular at low-depth circuits where it shows advantage over alternative approaches. The Feynman path sum technique we introduce here can also be adapted for the simulation of interference in certain nonlinear interferometers.

The paper is organized as follows.  In Section \ref{sec:fps}, we describe the process of computing a boson sampling probability amplitude using a Feynman path sum and tensor contraction. In Section \ref{sec:depth}, we verify the correctness of our algorithm implementation and apply it to the simulation of locally-connected interferometer designs featuring varying numbers of layers, modes, and photons. We benchmark it against other algorithms for the calculation of the permanent of matrices with specific characteristics, such as small treewidth. Finally, we provide some concluding remarks in Section \ref{sec:conclusion}.

\section{Feynman paths for the simulation of linear-optical interferometers} \label{sec:fps}

Boson sampling involves evolving an input Fock state through an $M$-mode interferometer and counting photons at the outputs using detectors, as shown in Fig. \ref{fig:U_inp_out}. The input state can be described by a tuple of occupation numbers $\ket{x} = \ket{x_1, ..., x_M}$, where $x_j$ is the number of photons in input mode $j$. The preserved total number of photons $N$ is obtained by summing occupation numbers over all modes: $N= \sum_{m=1}^{M} x_m$.

\begin{figure*}[ht!]
    \centering
    \includegraphics[width=0.8\textwidth]{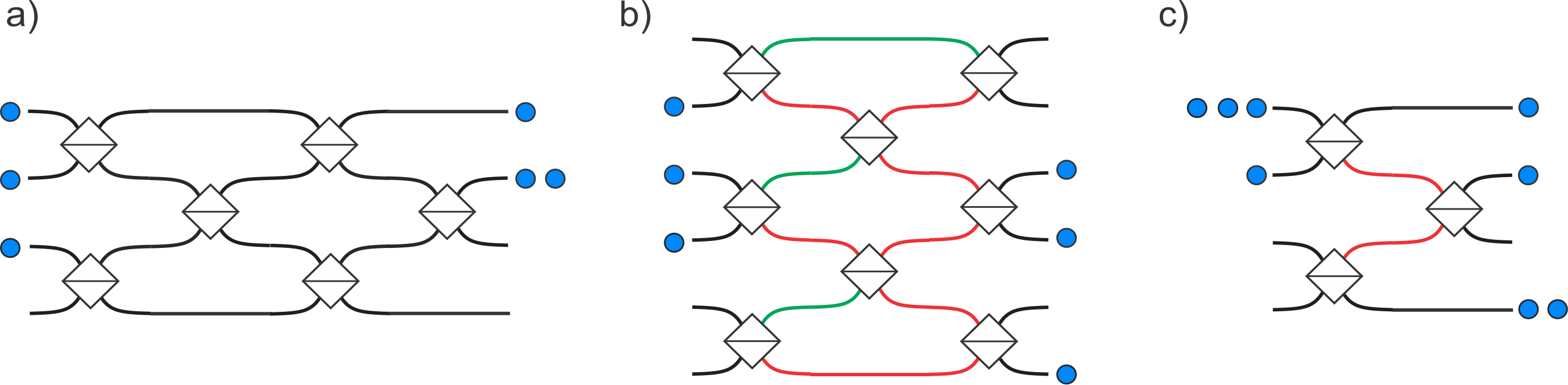} 
\caption{Feynman path simulation of Fock-state linear optics. (a) The interferometers are built from a mesh of locally-connected BS (squares). Blue circles represent target choices of photon number occupations at input and output modes. (b) The Feynman path simulation algorithm adds the probability amplitudes of all possible paths, each determined by allowed photon numbers at internal green and red waveguides. The path enumeration only needs to iterate over possible occupations of green-colored waveguides, as those of the red-colored ones are then uniquely determined due to photon number conservation. (c) We further simplify path enumeration by using light-cone reasoning to quickly identify invalid, zero-amplitude paths. As shown in this example, invalid paths can result from an incompatibility between the target input and output configurations and the interferometer connectivity.}
    \label{fig:exemplo}
\end{figure*}

A linear interferometer with $M$ modes can be decomposed into an array of $O(M^2)$ interconnected beam splitters (BS) and phase shifters using known decompositions ~\cite{ReckZBB94, Clements16}, see Fig. \ref{fig:exemplo}(a). We denote by $U$ the $M\times M$ unitary transformation mapping input to output creation operators.

The occupation numbers at the output are described by a tuple of integers $\ket{y} = \ket{y_1, ..., y_M}$. We are interested in evaluating the probability amplitude associated with the transition from a chosen input state to a chosen output state, given by $\bra{y}U_P\ket{x}$. Here, $U_P$ is the linear-optical unitary acting on the multi-photon Hilbert space and is completely defined by the unitary $U$.

\begin{figure}[ht!]
    \centering
    \includegraphics[width=0.45\textwidth]{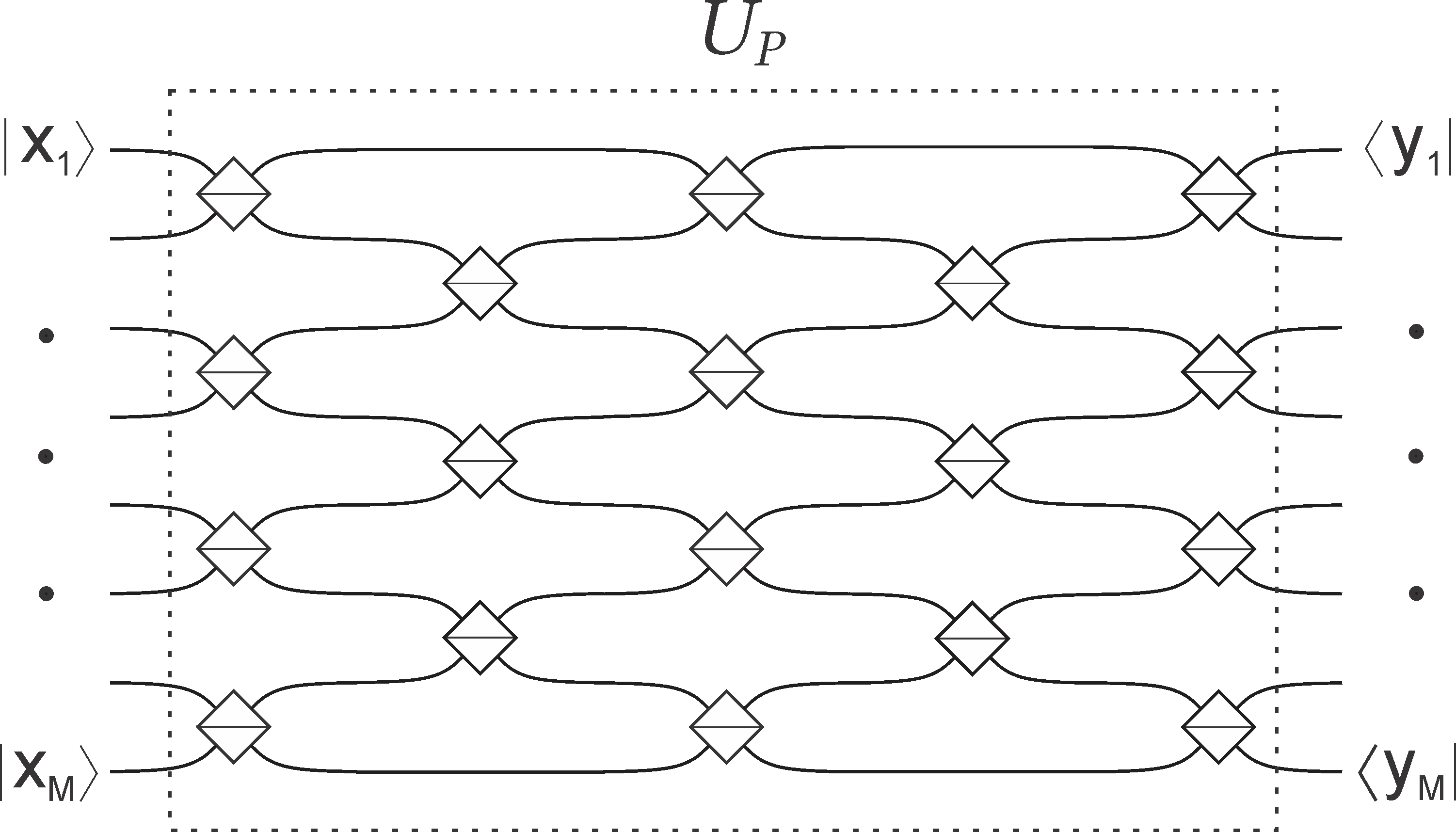}
    \caption{Boson sampling setup using a planar interferometer design consisting of locally connected BS, as in the universal design proposed by Clements \textit{et al.} \cite{Clements16}. Input photon occupations are denoted by $x_i$, and output occupations by $y_i$. 
    }
    \label{fig:U_inp_out}
\end{figure}

To express this amplitude as a Feynman path sum, we may decompose the interferometer into a network of basic linear-optical elements described by a $2 \times 2$ unitary matrix representing a phase shifter and a BS, jointly characterized by two parameters, $\theta$ and $\phi$ (where $\theta \in \left[0,\pi/2\right]$ and $\phi \in \left[0,\pi\right]$). $\theta$ determines the transmissivity of the BS and $\phi$ describes the phase difference between the interferometer arms, as described in Eq. \ref{eq:bs_matrix}. For convenience, we refer to this optical element simply as a (generalized) BS, described by:

\begin{equation}
BS = \begin{bmatrix} 
\cos{\theta} & -e^{-i \phi} \sin{\theta} \\ 
e^{i \phi} \sin{\theta}  & \cos{\theta}
\end{bmatrix}.
\label{eq:bs_matrix}
\end{equation}

Each BS is represented by a unitary map $BS_i$ from input to output creation operators, where $i$ denotes the label of each BS in the overall interferometer. Thus, the complex probability amplitude for each BS is defined as $\bra{y_{1i},y_{2i}} BS_i\ket{x_{1i}, x_{2i}}$, where $x_{1i}$ and $x_{2i}$ ($y_{1i}$ and $y_{2i}$) represent the number of input (output) photons in the upper (lower) modes in each BS$_i$. In Appendix \ref{sec:appendix}, we obtain a formula for BS amplitudes of that form, with an efficient time complexity scaling linearly with the number of photons $O(N)$. This routine may also be helpful for other methods involving simulation of physical systems using linear-optical interference of Fock states, as discussed for example in \cite{Stob19}.

To evaluate the probability amplitude of interest, we define a \textit{valid path} as a possible tuple of occupation numbers for all modes connecting BS, given the constraints imposed by the chosen occupation numbers at input and output. As examples of invalid paths, in Fig. \ref{fig:exemplo}(b) any non-zero occupation for the top green mode would be invalid; and in Fig. \ref{fig:exemplo}(c), there are no valid occupations for the red-colored paths.  For each valid path, we evaluate the probability amplitude in each BS, then multiply these individual amplitudes to obtain the probability amplitude associated with the path: $\prod_{i=1}^{N_{BS}}\bra{y_{1i},y_{2i}} BS_i\ket{x_{1i}, x_{2i}}$, where $N_{BS}$ is the total number of BS in the interferometer. Finally, by taking the Feynman sum over amplitudes for all valid paths $p$, we obtain the total probability amplitude for the boson sampling experiment:
\begin{equation}
    \bra{y}U_P\ket{x} = \sum_{p} \prod_{i=1}^{N_{BS}}  \bra{y_{1i}, y_{2i}} BS_i\ket{x_{1i},x_{2i}}.
    \label{eq:1}
\end{equation}

 A quick identification of invalid paths is crucial as it can significantly accelerate the computation by decreasing the number of amplitudes to be calculated. We also need a systematic way to enumerate all possible valid paths, whose associated amplitudes need to be added for the overall circuit amplitude calculation using Eq. (\ref{eq:1}). These important points, represented in Fig. \ref{fig:exemplo}(b)--(c), are discussed in detail in the next subsection.

\subsection{Amplitude calculation for low-depth circuits}

\begin{figure}[ht!]
    \centering
    \includegraphics[width=0.47\textwidth]{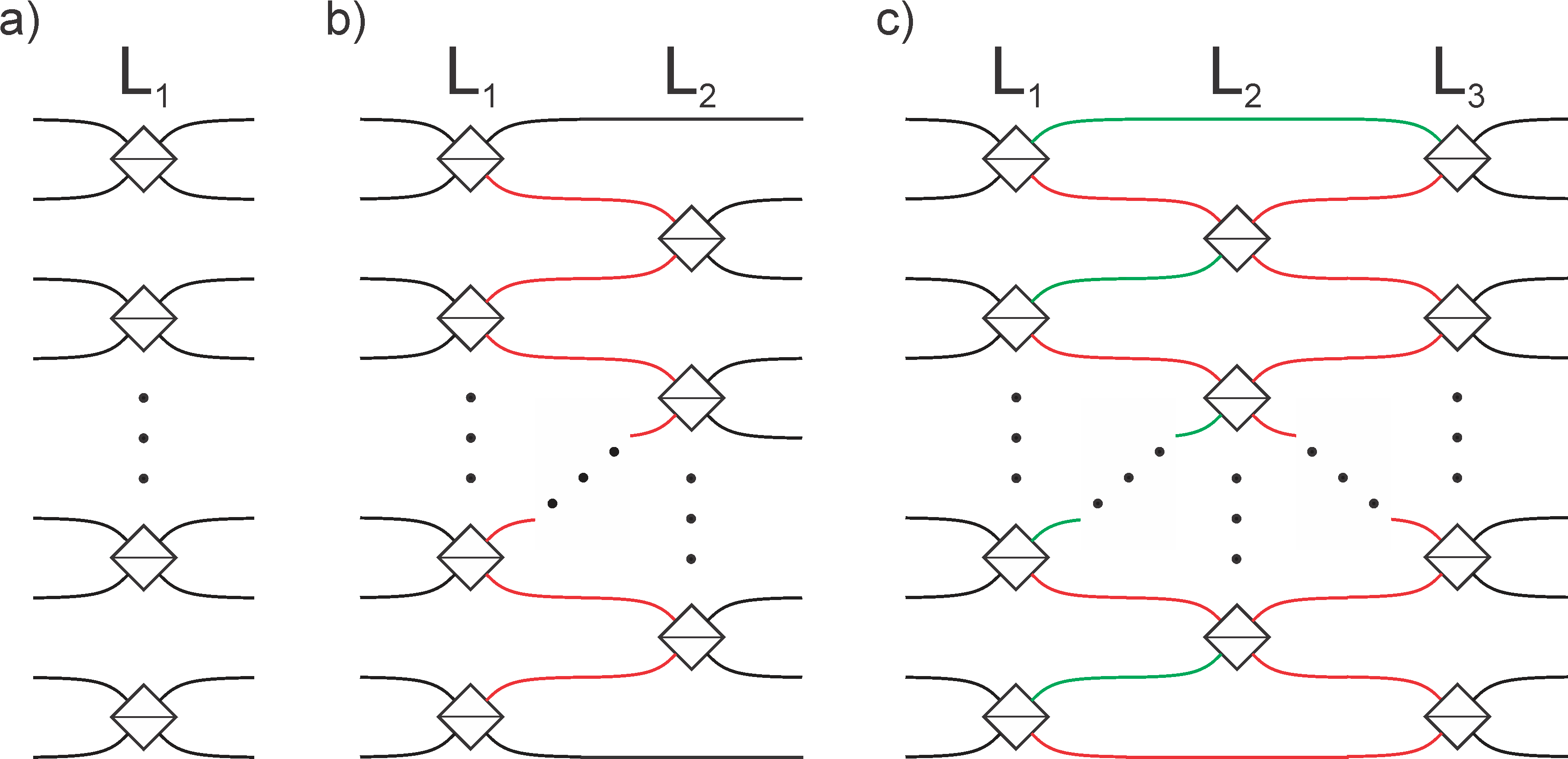}
    \caption{Planar interferometer design comprising (a) one, (b) two, and (c) three layers of locally-connected BS. The modes colored in black represent the target configurations of input and output. Modes colored green need to be iterated over for the Feynman path sum, while modes shown in red have their occupation numbers determined by those colored green or black, and the fixed input/output occupations, due to photon number conservation at each BS.
    }
    \label{fig:1_2_3_layers}
\end{figure}

We begin our analysis of the Feynman path simulation described above by considering shallow optical circuits with a planar design as in Fig. \ref{fig:U_inp_out}. To illustrate the method, Fig. \ref{fig:1_2_3_layers} displays circuits with increasing numbers of layers of BS. Note that, for an $M$-mode interferometer with this design, a minimum of $M$ BS layers are needed for a universal linear-optical device. The number of layers $D$ denotes the depth of the circuit. Figures \ref{fig:1_2_3_layers}(a)--(c) correspond to circuits with one, two, and three layers, respectively. We use a color scheme for the modes to help identify valid paths for the Feynman path calculation, as described in the caption.

The circuit with one BS layer of Fig. \ref{fig:1_2_3_layers}(a) has no connected paths to iterate over. In the language of Feynman paths, there is a single valid path, so the summation in Eq. (\ref{eq:1}) has only a single term given by the product of the probability amplitudes for each BS in this circuit. If the circuit has $M$ modes (assumed to be even), the calculation runtime scales linearly with $M$ as it corresponds to a straightforward multiplication of single BS amplitudes. In Appendix A we describe a simple algorithm for calculating each BS amplitude, having a runtime that is linear in the number of photons going through the BS. Consequently, for this one-layer, $M$-mode interferometer the overall runtime complexity increases linearly with both number of modes $M$ and total photon number $N$.

Fig. \ref{fig:1_2_3_layers}(b) shows a generic circuit with two BS layers. The BS connectivity, together with photon number conservation at each BS, allow us to unambiguously fix the occupation numbers of all waveguides connecting BS, depicted in red in the picture. In other words, for Fig. \ref{fig:1_2_3_layers}(a)--(b), if there is a valid path for the circuit, that path is unique. Therefore, the complexity remains linear with both the number of modes $M$ and the number of photons $N$. This will be the case even if the circuit design allows for random connectivity between distant BSs in the first and second layers.

Note that in the two types of interferometers we have just discussed, having only one or two BS layers means that certain outputs are impossible for a given input. This occurs when photon number conservation is violated at BS. This will happen for invalid paths that correspond to a null probability amplitude for the whole setup. Identifying such invalid paths is crucial for speeding up computation, as it allows us to discard these paths without the computational cost of evaluating all associated amplitudes.

In both cases described, Fig. \ref{fig:1_2_3_layers} (a)--(b), we have just one valid path. In this case, Eq. \ref{eq:1} simplifies to

\begin{equation}
    \bra{y}U_P\ket{x} = \prod_{i=1}^{N_{BS}}  \bra{y_{1i}, y_{2i}} BS_i\ket{x_{1i},x_{2i}}.
    \label{eq:10}
\end{equation}

Fig. \ref{fig:1_2_3_layers}(c) illustrates a circuit where the Feynman path summation of Eq. \ref{eq:1} is not as straightforward. In general, this allows for multiple possible configurations of photon number occupations for the modes connecting the BS layers. To enumerate all valid paths more efficiently, we will introduce a coloring of the modes into black, green, and red. As before, black waveguides are input and output modes whose occupation numbers are given. We now propose a way to color another subset of waveguides green, corresponding to modes whose occupation numbers are looped over in the sum over paths. Our choice of green waveguides is shown in Fig. \ref{fig:1_2_3_layers}(c), i.e. the top output mode of each BS in layer $L_1$. For each chosen set of occupation numbers for green waveguides, the remaining waveguides (in red) have their occupation numbers uniquely determined due to three factors: the fixed planar waveguide connectivity, the fixed input and output occupation numbers, and photon conservation at each BS.

To define the range of occupation numbers that need to be considered for each green waveguide, we must consider the minimum number of input photons ($N_i = x_{1i} + x_{2i}$) in each BS$_i$ of the first layer ($L_1$ in Fig. \ref{fig:1_2_3_layers}). This results in up to $N_i + 1$ possible output configurations for each BS$_i$, in one-to-one correspondence with the occupation number of each green waveguide. Therefore, the output occupation number of each BS in the first BS layer is defined as $[(y_{1i},y_{2i})] = [(N_i, 0), (N_i-1, 1), ..., (0, N_i)]$, corresponding to an iteration over all possible paths for each BS in the first layer ($L_1$). Note that $y_{1i}$ denotes the occupancy of green waveguide BS outputs (see Fig. \ref{fig:1_2_3_layers}(c), with $y_{2i}$ denoting the occupancy of BS output waveguides marked as red.

It is important to note that during the enumeration process over possible occupations of green waveguides, we may reach invalid paths, i.e. paths associated with a null probability amplitude. This is revealed by a mismatch between occupation numbers at the input and output of some BS, which reflects the impossibility of satisfying all photon number conservation rules for that particular configuration.

In summary, the Feynman path strategy to simulate depth-3 interferometers proceeds as follows: we start with the given input and output occupation numbers, an upper bound is obtained for the occupation numbers of the green waveguides, and this allows iteration over all possible configurations of green waveguide occupation numbers. These configurations, in turn, determine the occupation numbers for the red waveguides, as shown in Fig. \ref{fig:1_2_3_layers}(c). For each valid path, we calculate the probability amplitude by multiplying the amplitudes associated with each BS. By summing the amplitudes of all valid paths, we obtain the total probability amplitude, as in Eq. \ref{eq:1}.

 The number of green waveguides increases linearly with the number of modes $M$, leading to a runtime that increases exponentially with $M$. Additionally, an exponential runtime dependency on depth $D$ will arise as we scale up the depth, something we discuss in the next subsection.

\subsection{Simulating circuits of arbitrary depth}

\begin{figure}[ht!]
    \centering
    \includegraphics[width=0.47\textwidth]{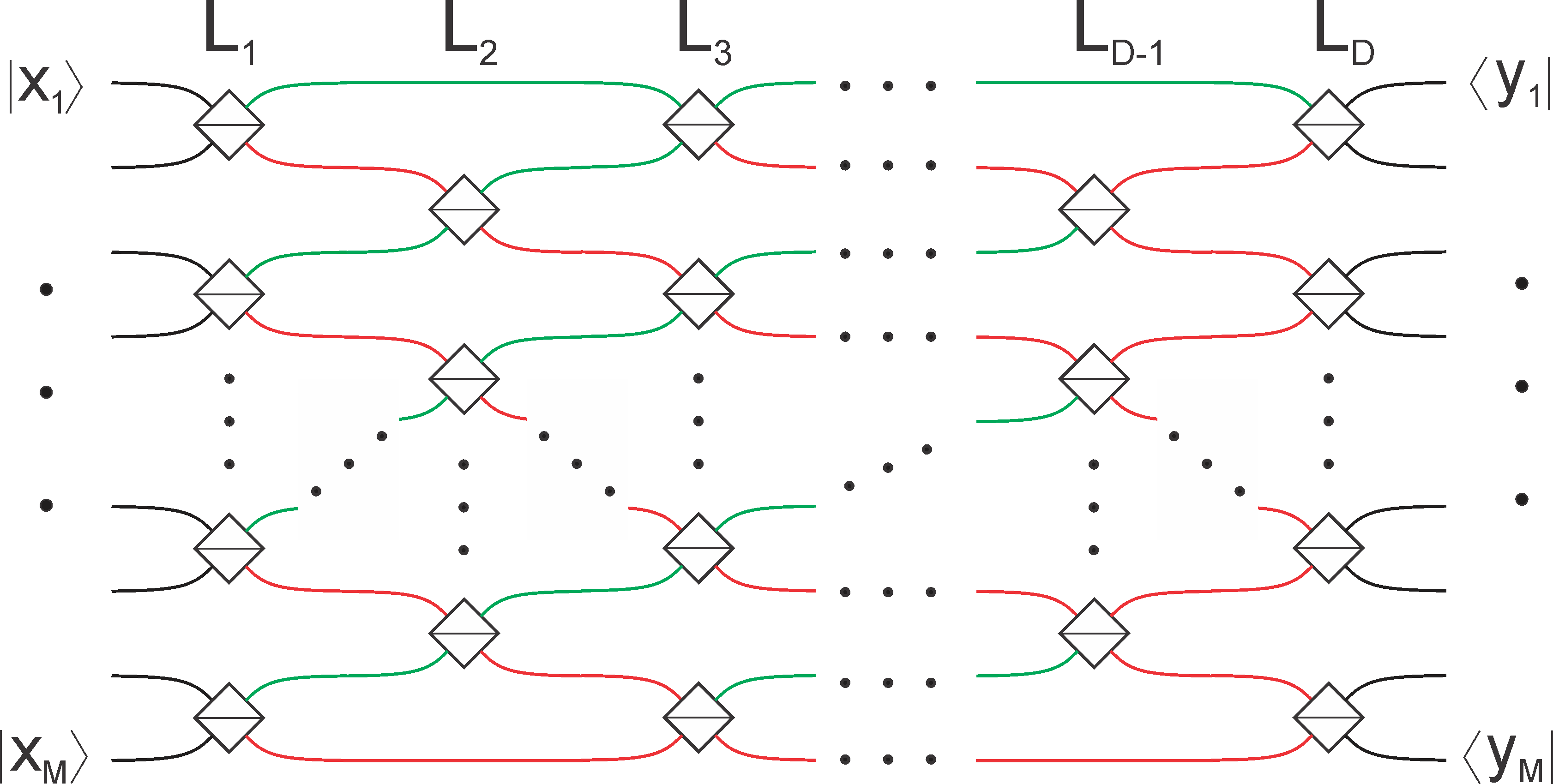}
    \caption{Planar interferometer design consisting of $D$ layers of locally-connected BS. A Feynman path amplitude calculation requires iterating over all occupation numbers for green-colored modes, which then uniquely determine those of the red-colored ones.} 
    \label{Fig2::general_circuit}
\end{figure}

Let us now consider how to apply the Feynman path simulation technique to a planar, locally-connected multimode interferometer with $D$ layers of BS and $M$ modes, as depicted in Fig. \ref{Fig2::general_circuit}. The procedure here follows the same steps described for $D=3$ in the last section, except we now also need green-colored modes -- whose occupation numbers define the occupation numbers of red-colored modes -- in all layers from $L_1$ to $L_{D-2}$. In Fig. \ref{Fig2::general_circuit}, note that photon number conservation at each BS, together with the fixed output occupation numbers, means we do not need green-colored waveguides connected to the last BS layer $L_{D}$ (with the exception of the top mode, which connects output $\ket{y_1}$ to a BS in layer $L_{D-2}$). The amplitude calculation using Feynman paths only iterates over the possible occupation numbers of green-colored waveguides, with the red-colored waveguide occupation numbers uniquely determined by those of the green ones, due to photon number conservation at each BS.

\begin{figure}[ht!]
    \centering
    \includegraphics[width=0.47\textwidth]{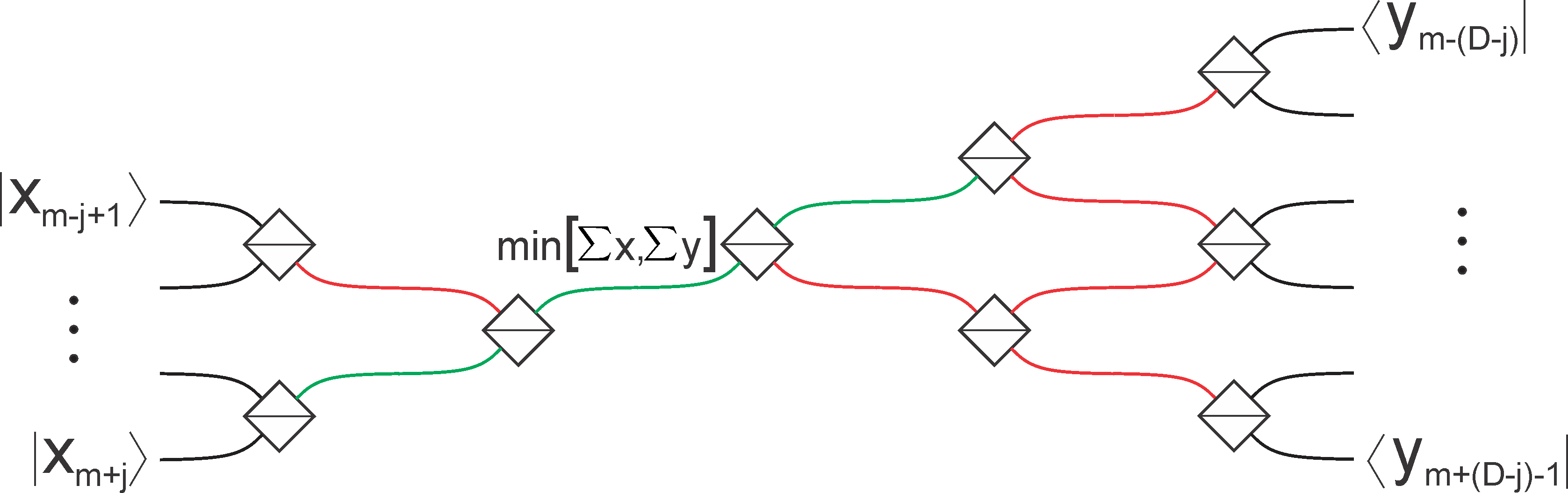}
    \caption{Past and future light cones for a green waveguide in the second BS layer of a 5-layer 
    interferometer. }
    \label{fig:light_cones}
\end{figure}

A strategy that simplifies the task of Feynman path enumeration is to establish an upper bound on the number of photons that need to be considered in each green-colored waveguide, taking into account the past and future light cones associated with each green waveguide. This helps eliminate invalid paths associated with null probability amplitudes. The past light cone for a specific green waveguide encompasses all the photons in the input modes in its past light cone. Conversely, the future light cone encompasses all the photons at the output modes that are connected to this particular waveguide. Fig.\ref{fig:light_cones} illustrates the connectivity of a green-colored waveguide in the second layer $L_2$, with the left representing the past light cone and the right representing the future light cone. It is important to note that we consider a layer to be a set of BS and their output waveguides. To describe the circuit operation, we use two index schemes. Index $m$ denotes the number of modes, ranging from $1$ to $M$, within a specific layer. Index $j$ refers to the layer itself, ranging from $0$ to $D$, with $0$ representing the input layer and $D$ representing the layer containing the output photons.

The highlighted green waveguide in Fig. \ref{fig:light_cones} will receive a maximum number of photons given by the sum of occupation numbers of all input modes in its past light cone: $\sum_{m = i}^{f} x_m$, where $i$ and $f$ denote the initial and final mode indices delimiting the light cone input. The highlighted green waveguide can only output up to  $\sum_{m = i}^{f} y_m$ photons in its future light cone. For each waveguide, we need to consider only the minimum value between the total numbers of photons in past and future cones, expressed as $\min\left[\sum_{m = i}^{f} x_m, \sum_{m = i}^{f} y_m\right]$. For example, suppose we have an eight-mode interferometer with five layers ($M=8,D = 5$). The green waveguide highlighted in Fig.\ref{fig:light_cones} is in position $m = 4$ and $j = 2$. In this case, we have $\ket{x_{m-j+1}} = \ket{x_{3}}$, $\ket{x_{m+j}} = \ket{x_{6}}$, $\ket{y_{m-(D-j)}} = \ket{y_{1}}$ and $\ket{y_{m+(D-j)-1}} = \ket{y_{6}}$, consequently the upper bound of the highlighted green waveguide is given by min$\left[\sum_{m = 3}^{6} x_m, \sum_{m = 1}^{6} y_m\right]$.

Once the occupation numbers of green waveguides are defined, we can proceed with the same methodology used for $D=3$: the input and output states (black waveguides) and all green waveguides, arranged in a nested loop, determine the occupations of red waveguides. Each such choice of possible occupation number is a valid path to be considered for the sum over paths, with the probability amplitude computed using Eq. \ref{eq:1}.
Unlike the case of depth-2 interferometers, for depth $D \ge 3$ there will be an exponential dependence of the runtime with both the number of modes $M$ and depth $D$ (if we assume a constant photon density $k=N/M$ at both input and output). This is because the number of green-colored waveguides increases linearly with $M$ and $D$, and the typical range of occupations numbers that need to be considered for each green waveguide increases with $k$.

The Feynman sum over paths method incurs minimal memory costs as the evaluation of the probability amplitude associated with each path scales polynomially with $M, D$, and one only needs to store the updated value of the sums and their indices as the computation proceeds.

\subsection{Feynman path simulation with tensor contractions}

\begin{figure}[ht]
\centering \includegraphics[scale=0.30,trim=0cm 0cm 0cm 0cm, clip=true]{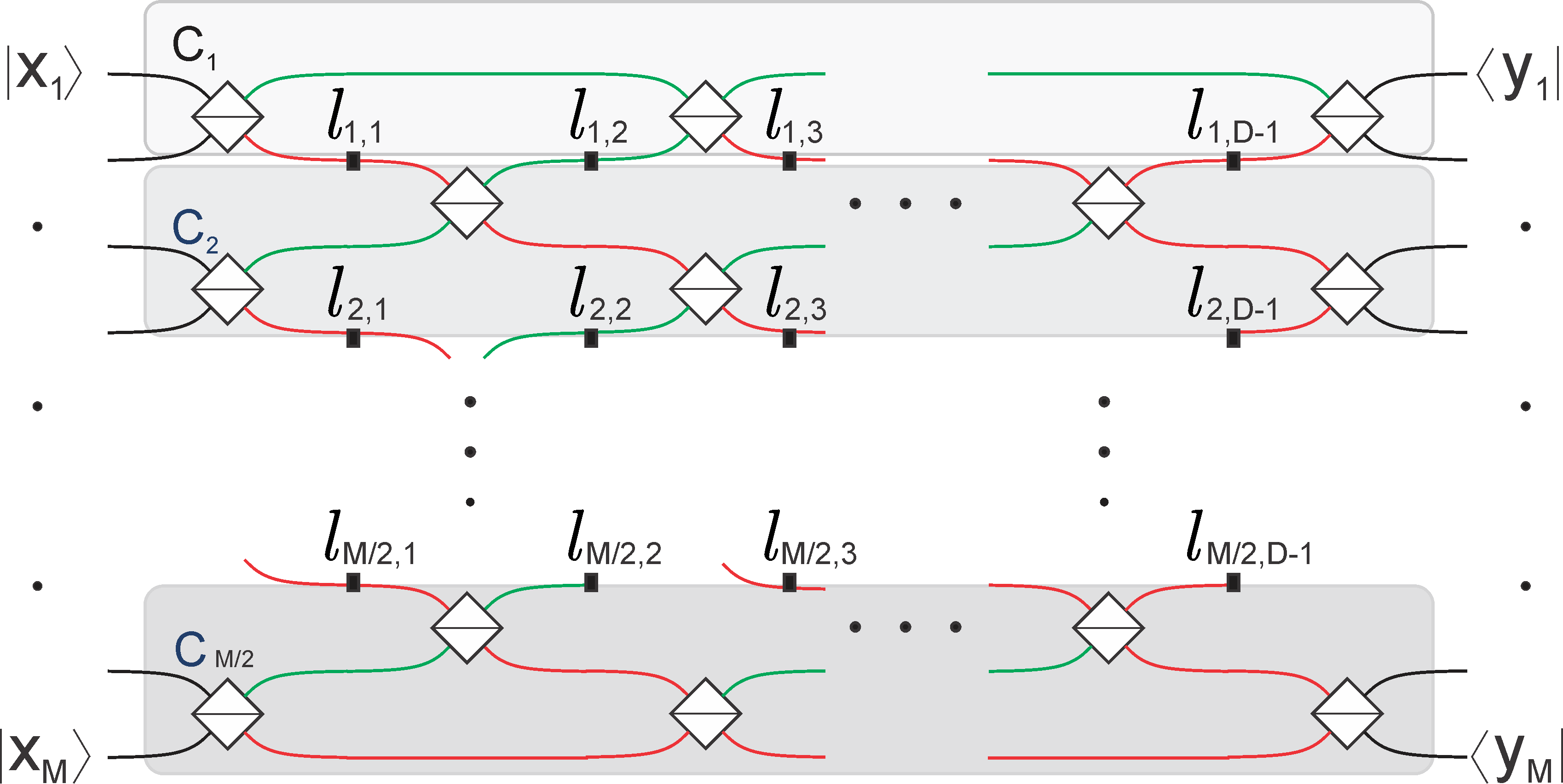}
\caption{Planar interferometer with tensor contraction layers $C_k$ highlighted in the alternating shaded boxes stacked vertically. $l_{k,j}$ denotes a tuple of photon occupation numbers and associated amplitudes for waveguides crossing between $C_k$ and $C_{k+1}$. Label $k$ ranges between $1$ and $M/2$, and $j$ ranges between $1$ and $D-1$. The tuple $l_{k,j}$ of occupation numbers and amplitudes stores all the information required to process the next layer, up until the last contraction layer, whose processing gives us the wanted amplitude. }
\label{contraction}
\end{figure}

The exponential runtime dependence on the number of modes (assuming constant photon number density per mode $k=N/M$) can be brought down to a linear dependence by using a tensor contraction technique ~\cite{MarkovS08}. As a result of the runtime improvement, a trade-off emerges, involving an exponential increase in the required memory. To achieve this, it is necessary to partition the circuit into $C_{M/2}$ contraction layers and then perform the Feynman path sum one layer at a time, as detailed below. The choice of tensor contraction order is arbitrary -- for example, one may contract the tensor from left to right, or from top to bottom. For the practical software implementation described in the next section, we have chosen to process the contraction blocks from top to bottom of the interferometer, with two modes per contraction block. As we will see later, this enables a simulation with a runtime that scales linearly with the number of modes, suitable for shallow-depth interferometers. Figure \ref{contraction} illustrates our choice of contraction blocks $C_{k}$ for the locally-connected 
interferometers we have been considering, with each horizontal block $C_{k}$ delineated by dark blue lines. The label $k$ ranges from $1$ to $M/2$, as each contraction block involves only two modes.

Let us begin by processing the first contraction layer, $C_{1}$. The initial step involves defining the range of occupation numbers to be considered for each green-colored waveguide in $C_{1}$. Convenient upper bounds for these occupation numbers are obtained using the past and future light cones, as discussed earlier. Unlike the path sum approach described in the last section, the tensor contraction will require storing intermediate occupation numbers for waveguides in between blocks, together with their associated amplitudes. More precisely, now we will do the sum over green-colored waveguides, taking note of all possible tuples of occupation numbers for waveguides that cross the border between the contraction layers $C_1$ and $C_2$ (denoted $l_{1,1}, l_{1,2}, ..., l_{1,D-1}$ in Fig. \ref{contraction}). As before, the input/output photons and the occupation numbers of the green waveguides determine the occupation of the red waveguides. By looping over all possible values for the occupation numbers of green waveguides, we obtain and store one probability amplitude associated with each possible value of the tuple $l_{1,j}$ of occupation numbers in the border between $C_1$ and $C_2$, using Eq. \ref{eq:1}.

The tensor contraction corresponding to each following contraction block $C_t$ works similarly: we loop over all possible stored occupation numbers corresponding to the previous block $C_{t-1}$, each with its associated probability amplitude. For each value of the tuple, we then loop over all possible values for the green waveguides within $C_t$, to obtain and store a tuple $l_{t,j}$ of possible occupation numbers for waveguides crossing from $C_t$ to $C_{t+1}$, and their associated probability amplitude (which is multiplied by the initial amplitude stored together with the input list $l_{t-1,j}$).

Processing the last contraction block $C_{M/2}$ is simpler, as the sum over the green waveguide occupation numbers, together with the stored input amplitudes, will directly give us the probability amplitude of the whole circuit.

Note that this approach can lead to significant memory usage as during the computation corresponding to block $C_t$, one needs to have access to the stored ingoing tuple of occupation numbers $l_{t-1, i}$, together with their associated amplitudes. We also need to store the list  $l_{t, i}$ of outgoing occupation numbers, together with their associated amplitudes. This results in both runtime and memory that scale exponentially with depth $D$, due to the exponential increase in size of the stored lists of partial amplitudes. We will highlight these practical limitations in the next section, where we discuss the performance of a software implementation of this tensor contraction technique.

\section{Numerical simulation results and comparison with other algorithms} \label{sec:depth}

In this section, we evaluate the efficiency of the Feynman path approach to the calculation of probability amplitudes of linear-optical experiments. We will consider the problem of simulating a rectangular architecture as depicted in Fig. \ref{Fig2::general_circuit}, with varying depth, number of modes, and number of photons. All analyses are supported by robust simulations of multimode interferometers using the Feynman path technique, implemented as a Linear-Optical Feynman Paths (LOFP) software package in C, available in the Zenodo repository at \href{https://zenodo.org/records/7681675}{https://zenodo.org/records/7681675}. In this section we benchmark this software by comparing it with other algorithms for strong simulation of linear optics: Ryser \cite{Ryser1963} and Cifuentes-Parrilo \cite{CifuentesP16}, which are also implemented in the same Zenodo repository. Additionally, the code implements the tensor contraction technique, which significantly enhances the performance of our LOPF simulation.

The tensor contraction technique is implemented for all input circuits of depth $D>2$. For $1$- and $2$-depth 
interferometer structures, all simulations have linear runtime, making the tensor contraction technique unnecessary. For Ryser's formula, the implementation is optimized using Gray code order \cite{gray1953pulse}, which achieves an asymptotic runtime complexity of $\mathcal{O}(2^{n}n)$, where $n$ is the number of photons. The Cifuentes-Parrilo algorithm \cite{CifuentesP16} exploits the sparseness structure arising from small-depth circuits,  whose simulation involves the calculation of the permanent of matrices with small treewidth $\omega$. Its complexity is $\mathcal{\tilde{O}}(n4^{\omega})$, where $\mathcal{\tilde{O}}$ denotes complexity ignoring polynomial factors in $\omega$. All algorithms implemented are currently nonparallelized.

All calculations were done using a laptop computer with the following specifications: Processor – Intel(R) Core(TM) i7-8565U, CPU @ 1.80GHzb – 1.99 GHz; Installed RAM: 16.0 GB (15.8 GB usable); System type: 64-bit operating system, x64-based processor.
 
\subsection{Verifying correctness}

In this subsection, we address the reliability of the code. To this end, we choose to simulate different input states on a $6$--mode interferometer. For each input state and a fixed interferometer, we use the Linear-Optical Feynman path (LOFP), Ryser, and Cifuentes-Parrilo algorithms to evaluate the output probabilities for all possible outputs. Ideally, the sum of all probabilities should be correctly normalised. 
We compare the evaluated probability of LOFP vs. Ryser and LOFP vs. Cifuentes-Parrilo using the total variation distance (TVD) between the calculated output distributions of pairs of methods $p,q$:

\begin{equation}
\delta(p, q) = \frac{1}{2} \sum_i |p_i - q_i|,
\end{equation}
\noindent where $p_i$ are the output probabilities obtained with our LOFP simulator, and $q_i$ are the probabilities calculated using either Ryser  or Cifuentes-Parrilo algorithms. 

We run the test using four different input states and depths: $\ket{1, 1, 1, 1, 1, 1}$ ($D = 3$), $\ket{0, 0, 4, 0, 0, 4}$ ($D = 4$), $\ket{2, 0, 3, 0, 0, 3}$ ($D = 5$), $\ket{2, 0, 0, 2, 0, 2}$ ($D = 6$). For each input state, we execute the code while considering all possible outputs to evaluate its reliability. In all cases, the probabilities of all output events sum to 1 within numerical accuracy, as expected. In all four cases, the code demonstrates high accuracy, achieving a TVD of $10^{-13}$ for LOFP and Ryser, and for the pair LOFP and Cifuentes-Parrilo, the numerical accuracy is of the order of $10^{-16}$.

\subsection{Interferometers with two beam splitter layers}

Let us start our analysis with an interesting distinct case: depth-2
interferometers as depicted in Fig. \ref{fig:1_2_3_layers}(b). Here, we compare our LOFP algorithm (Eq. \ref{eq:1}) with the algorithms proposed by Ryser \cite{Ryser1963}, and Cifuentes and Parrilo \cite{CifuentesP16}. We selected Ryser's algorithm for comparison because it is known to be faster in scenarios with a low number of photons (less than $\sim 30$). Conversely, the Cifuentes-Parrilo algorithm demonstrates an advantage in scenarios involving the calculation of permanents of matrices with small bandwidth, 
which is the case of depth-2 interferometers.

\begin{figure}[h!]
    \centering
    \includegraphics[scale=0.50,trim=0cm 0cm 0cm 0cm, clip=true]{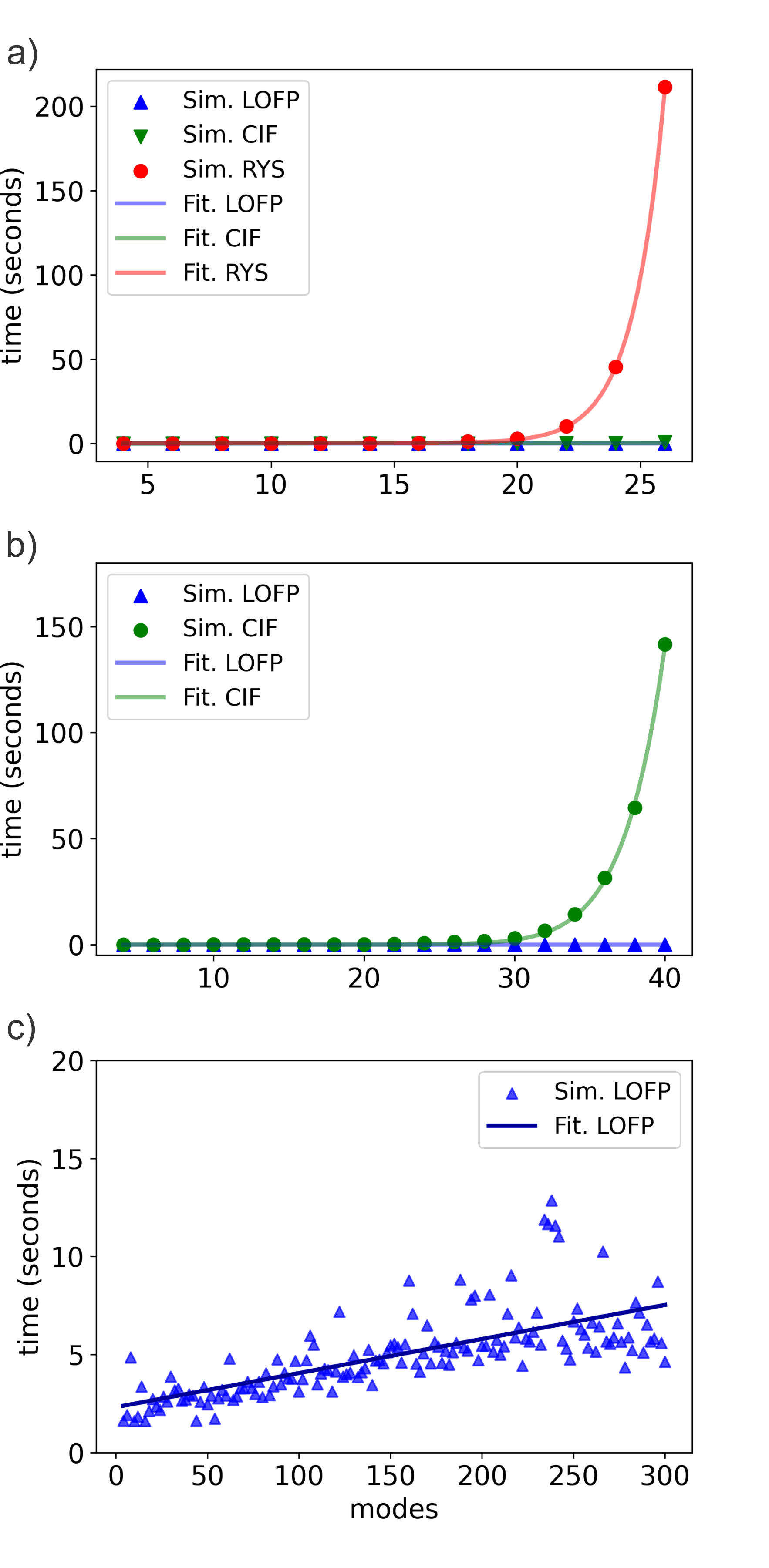}
    \caption{Runtime comparison between the linear-optical Feynman path (LOFP), Ryser, and Cifuentes-Parrilo algorithms for simulation of $2$-depth interferometers. In all cases, we consider even numbers of modes. Subfigures a,b have one photon per input and output mode, subfigure C has five photons per input/output mode. Comparison, as a function of number of modes, of runtimes: (a) of the three algorithms for low numbers of modes; (b) LOFP and Ryser algorithms for higher number of modes;
    (c) LOFP simulation with up to $300$ modes and $1500$ photons.  }
\label{fig:2layers}
\end{figure}

Fig. \ref{fig:2layers}(a) shows the runtime comparison between our Linear-Optical Feynman Path (LOFP) algorithm, Ryser's algorithm, and the Cifuentes-Parrilo algorithm. In Fig. \ref{fig:2layers}(a), we see the algorithms' runtimes are very similar up to $\sim 20$ modes, at which point the exponential runtime scaling of Ryser's formula becomes apparent. To further compare the LOFP algorithm with the one by Cifuentes-Parrilo, we increased the number of modes/photons up to 40, as shown in Fig. \ref{fig:2layers}(b). Here, Cifuentes-Parrilo exhibits a polynomial runtime fitted with a power-law function, while Feynman path shows a linear runtime with the increasing number of modes (as expected from our analysis in section \ref{sec:fps}). Fig. \ref{fig:2layers}(c) illustrates the Feynman path runtime for 5 photons per input and output mode, and for up to 300 modes (i.e. up to 1500 photons), which is practically beyond what is possible with the algorithms of Cifuentes-Parrilo or Ryser.

As discussed in Sec.\ref{sec:fps}, for these depth-2 interferometers the Feynman path runtime complexity scales linearly with the number of modes and density of photons, and is much faster than the other algorithms.

\subsection{Deeper interferometers}

\subsubsection{LOFP linear time scaling with the number of modes}

\begin{figure}[h!]
    \centering
    \includegraphics[scale=0.58,trim=0cm 0cm 0cm 0cm, clip=true]{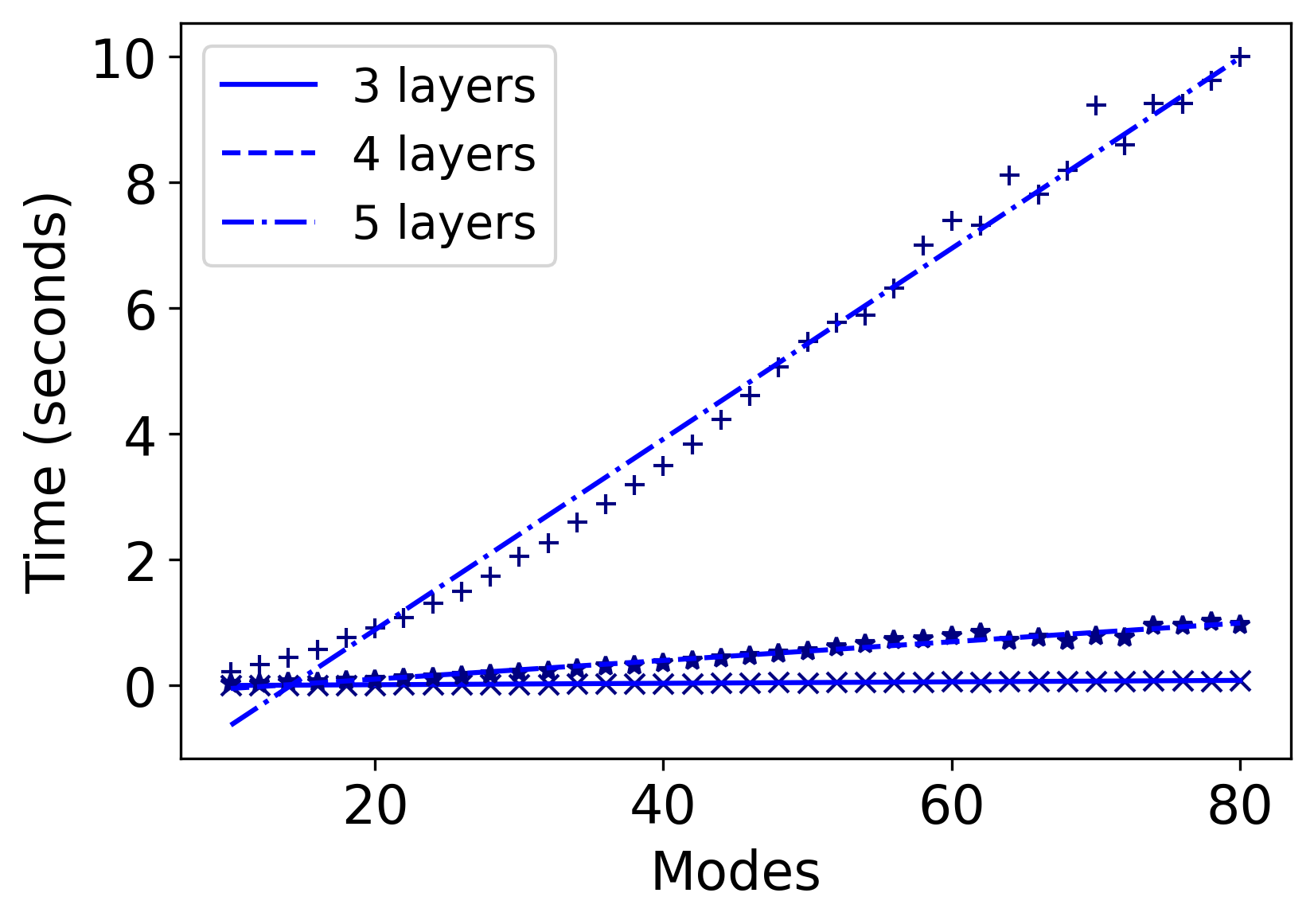} \caption{Runtime for linear-optical Feynman path (LOFP) simulation for interferometers with depth ranging from three to six, always with one photon per mode at each input and output mode.}
\label{Fig:linear_time}
\end{figure}

In this section, we simulate multimode interferometers with between three and five layers of BS, using our LOFP code. We fix the inputs and outputs to consist of a single photon per mode. 
The parameters $\theta, \phi$ describing each BS are sampled uniformly (see Eq. \ref{eq:bs_matrix}). 
The simulation results are presented in Fig.~\ref{Fig:linear_time}.

For the 3--5-layer interferometers we simulated, the Feynman path algorithm exhibits linear time scaling with respect to the number of modes/photons, due to the tensor contraction technique applied to these low-depth interferometers. Note that we only show runtimes starting from $20$-mode interferometers to ensure that the lists of amplitudes calculated by each contraction layer are roughly of the same size. This is important because, initially, these lists are shorter due to the interferometer boundary conditions, resulting in faster processing time for the first modes. These simulations show a significant advantage compared to Ryser's and Cifuentes' formulas, as those have a runtime that increases exponentially with the number of photons and modes. It is also evident that the linear coefficient of the LOFP code runtime increases with the number of layers. This occurs because the number of possible paths to be evaluated with each tensor contraction step increases exponentially with depth, as discussed in Section \ref{sec:fps} C.

\subsubsection{Exponential scaling of memory and runtime with depth}

\begin{figure}[ht!]
    \centering
    \includegraphics[scale=0.38,trim=0cm 0cm 0cm 0cm, clip=true]{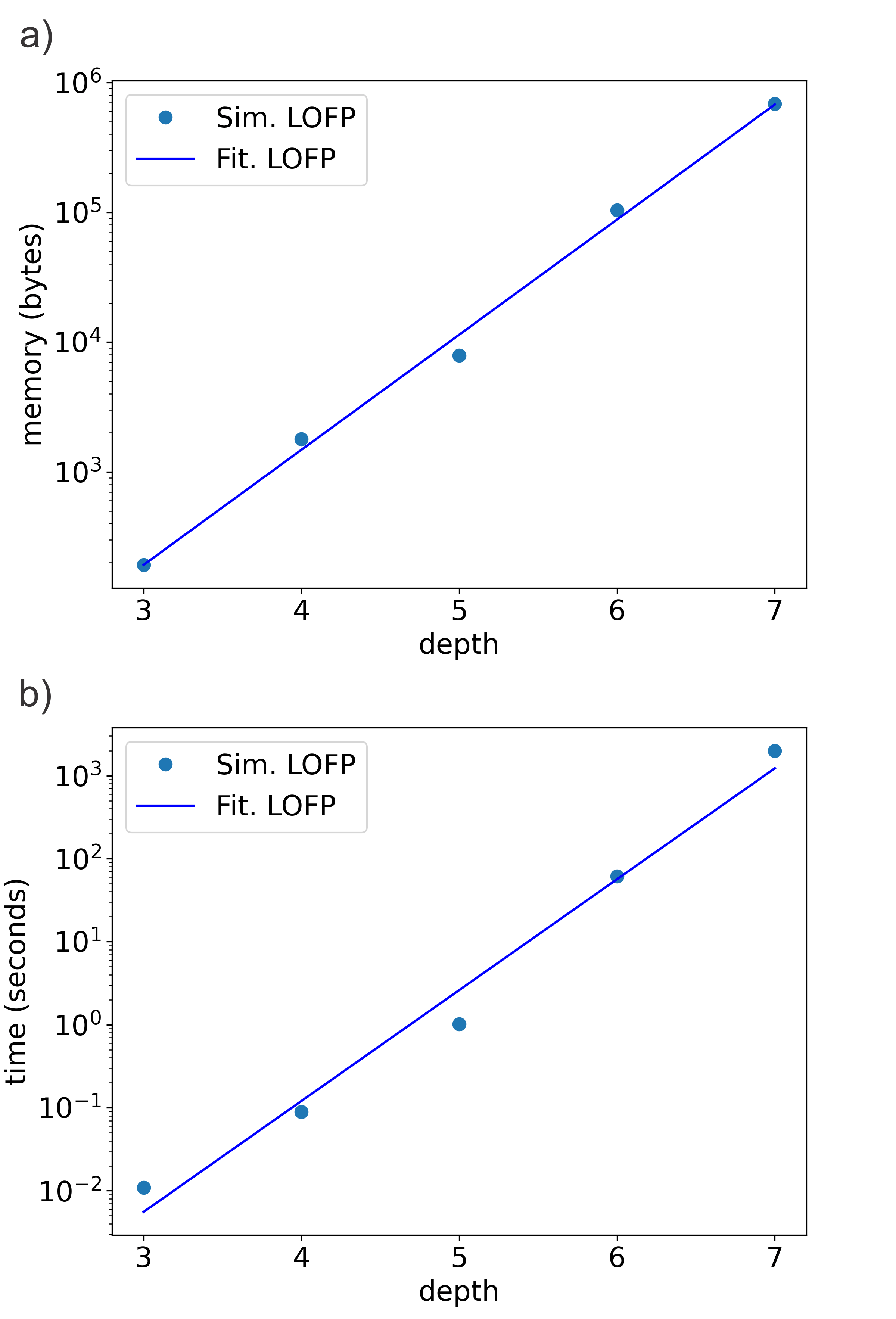}
    \caption{(a) Memory scaling for LOFP calculation of a single amplitude for $20$-mode interferometers with depth ranging from three to seven, with one photon per input and output mode. The data are plotted using a log-linear scale. (b) Runtime of the same simulation as a function of depth. The data are in a log-linear plot.}
\label{Fig:exponential_memory}
\end{figure}

This subsection discusses the memory and runtime requirements for our LOFP simulations using tensor contraction. Both memory and runtime scale exponentially with depth, due to the exponentially increasing number of inputs and outputs of each tensor contraction block that needs to be processed. We analyze simulations of 20-mode interferometers with increasing depth, considering 1 photon per mode at both input and output. The results are presented in Fig.~\ref{Fig:exponential_memory}.

As discussed in Section \ref{sec:fps}C, the result in Fig. \ref{Fig:exponential_memory}(a) shows that LOFP memory increases exponentially with depth when the tensor contraction technique is implemented. Figure \ref{Fig:exponential_memory}(a) also shows the highest memory usage among all contraction layers for each circuit instance. This peak memory usage occurs because we need to store the intermediate results from each layer during the contraction process. The highest memory usage appears for 7-layer interferometers, but even so, it remains relatively low, around a few Megabytes. It is clear that the memory footprint should be considered for larger simulations.

 Fig.~\ref{Fig:exponential_memory}(b) presents the simulation results, which illustrate the exponential scaling of the runtime with depth. This scaling goes hand-in-hand with the exponential memory scaling, as the memory is used to store inputs and outputs that need to be processed in each contraction layer. These results are consistent with the theoretical prediction outlined in Section~\ref{sec:fps}.C.

\subsubsection{LOFP versus Ryser's algorithm}

\begin{figure}[ht!]
    \includegraphics[scale=0.310,trim=0cm 0cm 0cm 0cm, clip=true]{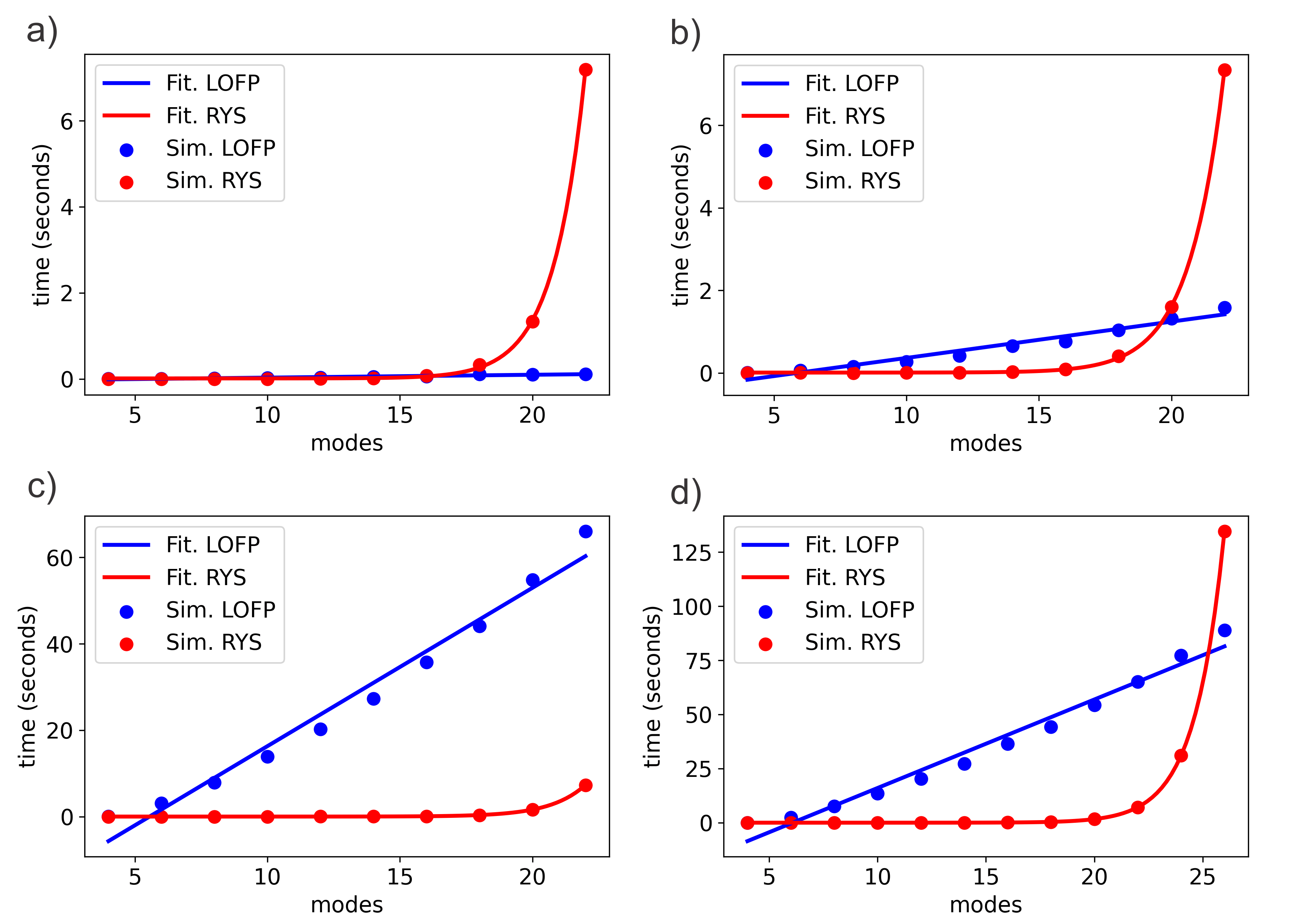}
    \caption{runtime for amplitude calculation of interferometers with depth $D$ ranging from four to six. The number of modes is increasing from $4$ to $22$, always with $1$ photon per mode. (a) $D = 4$, (b) $D = 5$, (c) and $D = 6$. In (d) $D = 6 $, and the number of modes increases from $4$ to $26$. Comparison between calculation time using Feynman paths (LOFP) and Ryser's algorithm with Gray code order (RYS).}
\label{fig:LOFP_Ryser}
\end{figure}

This subsection presents a comparative analysis of our Feynman path algorithm, which utilizes tensor contraction and Ryser's formulas enhanced with Gray code, focusing on their runtime performance. We simulate interferometers with an increasing number of modes and depth, featuring one photon per input and output mode. Fig.~\ref{fig:LOFP_Ryser} presents the runtime results for each algorithm across different depths, with subplots for (a) D = 4, (b) D = 5, and (c) D = 6.

Figure~\ref{fig:LOFP_Ryser} compares the runtime performance of our LOFP algorithm and Ryser's formula for different scenarios. The LOFP code outperforms Ryser for $D=4$ in Fig.~\ref{fig:LOFP_Ryser} (a). However, for $D = 5$ (Fig.~\ref{fig:LOFP_Ryser} (b), Ryser's formula is faster for a small number of photons. This advantage diminishes as the number of photons increases to 20, when LOFP shows an advantage. Similarly, in Fig.~\ref{fig:LOFP_Ryser} (c), for $D=6$, Ryser's initially exhibits faster runtime until the number of photons reaches $20$. 

Despite these cases where Ryser might seem favorable, Ryser's formula suffers from exponential runtime complexity with the number of photons. This becomes evident in Fig.~\ref{fig:LOFP_Ryser} (d), where the Feynman path calculation consistently shows faster runtime compared to Ryser for scenarios exceeding 26 modes/photons.

In conclusion, our linear-optical Feynman path (LOFP) algorithm is advantageous for small-depth interferometers and a large number of photons. It retains linear scaling with the number of modes and photon density at the input, as long as the memory is sufficient for the tensor contraction technique to be used.

\subsubsection{LOFP versus the Cifuentes-Parrilo algorithm}

\begin{figure}[ht!]
  \centering
  \includegraphics[width=0.9\linewidth]{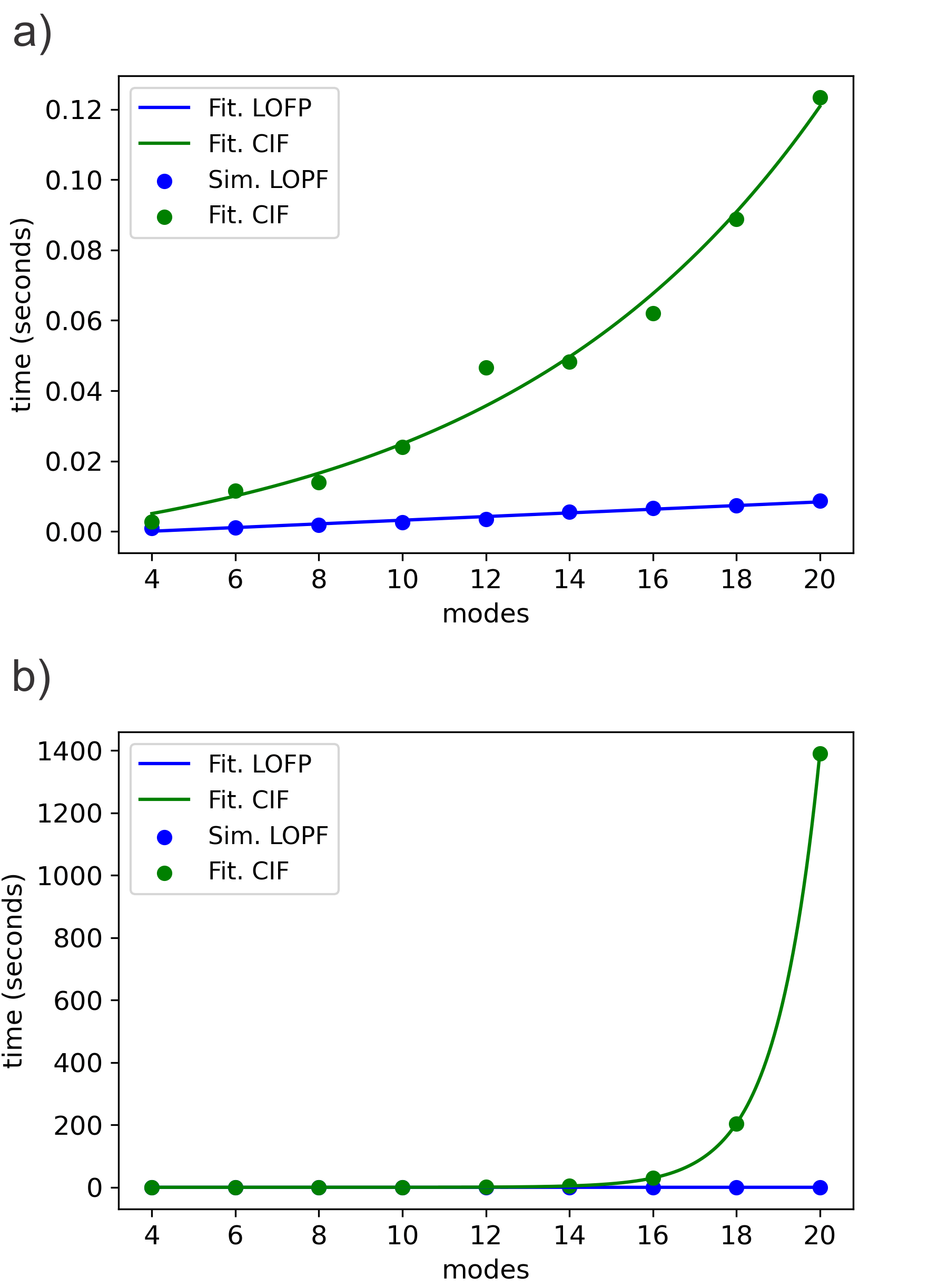} \\
  \caption{Amplitude calculation runtime for depth-3 (a) and depth-4 (b) interferometers. We always consider inputs and outputs with $1$ photon per mode. Comparison between calculation time using Feynman paths (LOFP) and Cifuentes-Parrilo (CIF).}
  \label{fig:combined_CIF_LOFP_1}
\end{figure}

In this section, we compare our LOFP algorithm's performance with the Cifuentes-Parrilo algorithm, briefly reviewed in section \ref{sec:depth}. In Fig. \ref{fig:combined_CIF_LOFP_1} we report amplitude calculations for interferometers with
an increasing number of modes, and with a single photon per input and output mode. The depth range from 3 to 4. We choose small depths because the Cifuentes-Parrilo algorithm is faster for permanent calculations of matrices with small treewidth $\omega$, as is the case for low-depth interferometers. Fig. \ref{fig:combined_CIF_LOFP_1} shows the comparison.

For small-depth interferometers, the Cifuentes-Parrilo algorithm may outperform LOFP for small numbers of modes or low photon densities at the input and output. However, as the number of photons and modes scales, the LOFP algorithm maintains linear behavior, while the Cifuentes-Parrilo algorithm exhibits polynomial growth.
Considering the exponential scaling of the Cifuentes-Parrilo algorithm with treewidth $\omega$, we expect exponential growth in scenarios where either depth or the photon density increases. The first one can be observed in Fig. \ref{fig:combined_CIF_LOFP_1}, where the runtime increases significantly when we go from depth-3 to depth-4 interferometers. This behavior is expected for both the Cifuentes-Parrilo and Feynman path approaches, whose runtime increases exponentially with depth.

\begin{figure}[ht!]
  \centering
  \includegraphics[width=0.8\linewidth]{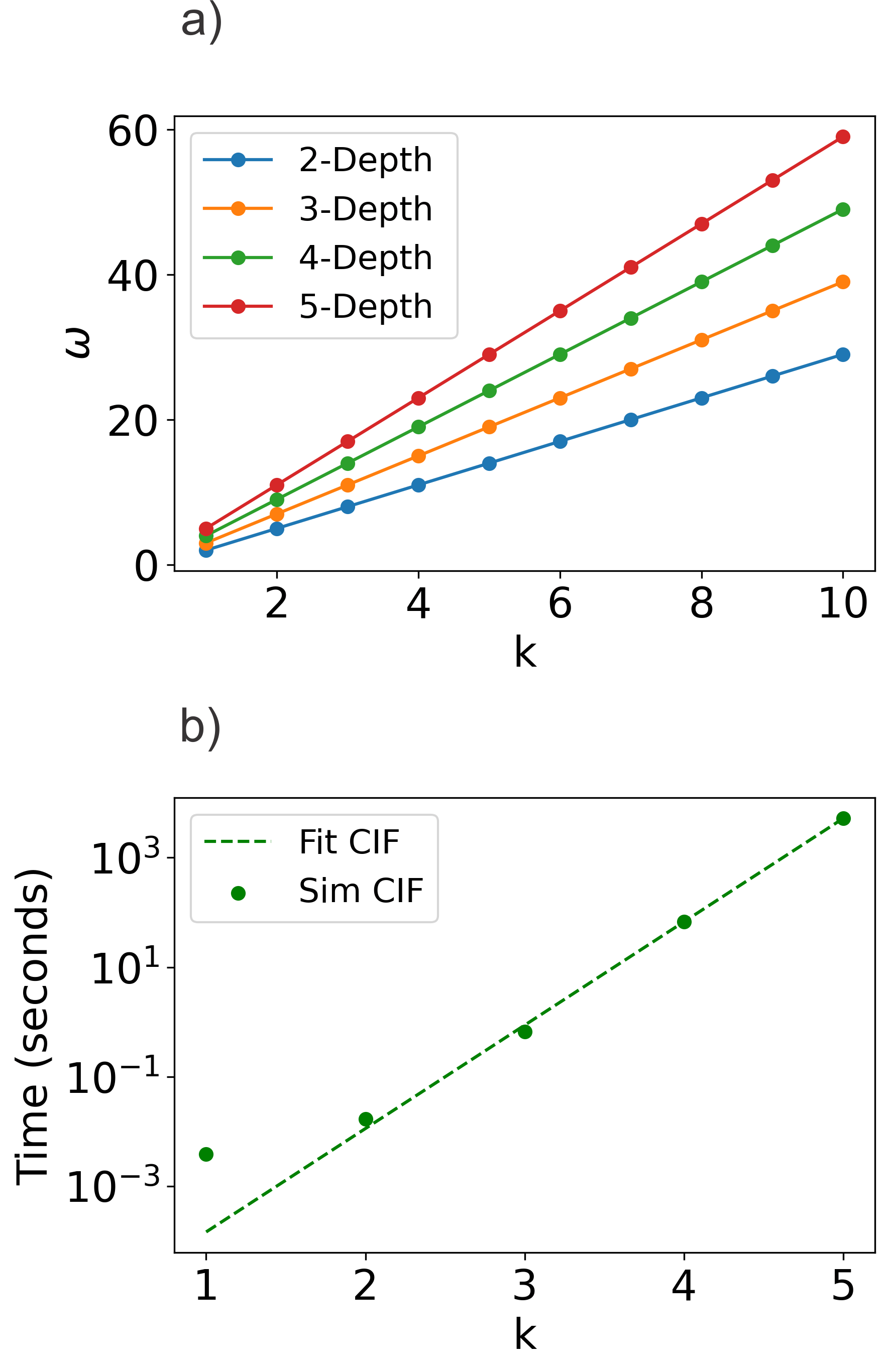} \\
  
  \caption{(a) shows the bandwidth $\omega$ behavior of the matrix whose permanent is calculated using the Cifuentes-Parrilo algorithm, as the photon density $k$ increases. Figure (b) is a log-linear plot of Cifuentes-Parrilo runtimes for depth-2 interferometers with $6$ modes, analyzed as a function of the photon density $k$ at the input and output. }
  \label{fig:combined_CIF_LOFP_2}
\end{figure}

Fig. \ref{fig:combined_CIF_LOFP_2}(a) shows how increasing the photon density $k$ affects the bandwidth $\omega$ of the matrix whose permanent calculation gives us the amplitude using the Cifuentes-Parrilo algorithm. We see the bandwidth $w$ scales linearly with the photon density $k$ at input and output. Since we know the runtime is exponential with $\omega$, we expect this algorithm will have a runtime that is exponential with $k$, which we confirm with the runtime data plotted in Fig. \ref{fig:combined_CIF_LOFP_2}(b). In contrast, in the next section, we will analyze the polynomial scaling of LOFP with increasing photon density.

\subsection{LOFP performance with higher numbers of photons at input and output}

\begin{figure}[ht!]
    \centering
    \includegraphics[scale=0.58,trim=0cm 0cm 0cm 0cm, clip=true]{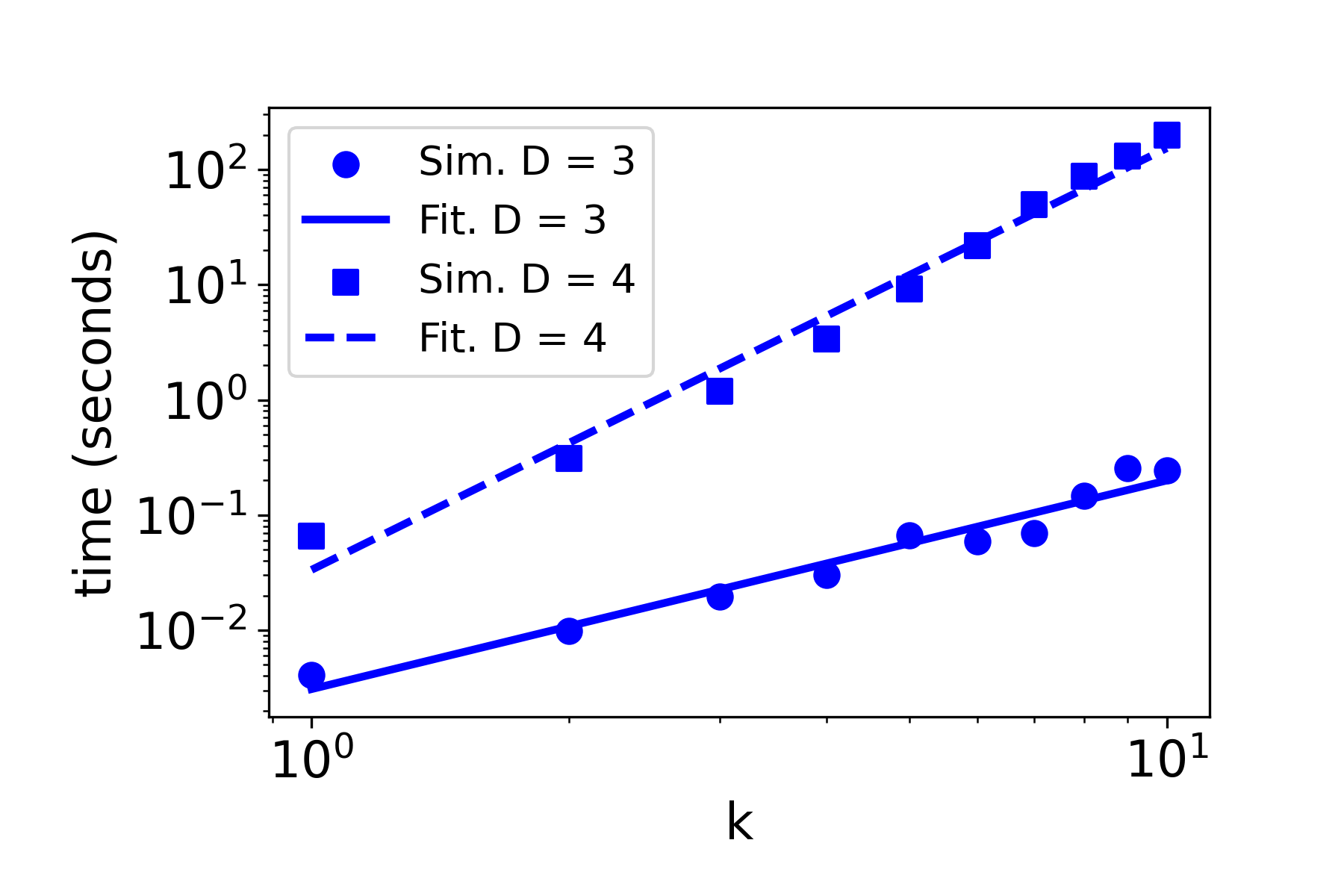}
    \caption{LOFP runtime for interferometer depth 3 and 4, with $10$ modes, as a function of photon density $k$ at input and output. The density of photons ($k$) increases from $1$ to $10$. The log-log plot indicates a polynomial scaling of the runtime with density of photons, as expected from our discussion in section \ref{sec:depth}.}
    \label{Fig:Higher_photons}
\end{figure}

Let us now study the LOFP performance for scenarios involving higher photon numbers at input and output. The simulation was performed for $10$-mode interferometers with depths $3$ and $4$. We start with $10$ photons ($1$ per mode) and add $1$ photon per mode in each step. So, the total number of photons varies from $10$ to $100$, in steps of $10$. The input and output states are always the same. Fig. \ref{Fig:Higher_photons} shows the results.

In this case, we can see the runtime scaling polynomially as the number of photons per input and output increases, as expected from our discussion in Section \ref{sec:fps}.C.

\section{Conclusion} \label{sec:conclusion}

We have described how to use Feynman's sum over paths technique for the calculation of probability amplitudes associated with boson sampling experiments, featuring Fock state input states and multimode linear-optical interferometers with photon-counting detection at the output. We have discussed specific strategies for speeding up the calculation for planar interferometer designs, in particular the identification of paths which do not contribute to the sum due to light-cone constraints and photon number conservation at each BS.

We showed how a tensor contraction technique reduces the runtime to a linear scaling with the number of modes and photons, assuming a constant density of photons per mode at input and output. This is done at the expense of exponential memory use, which nevertheless remains reasonably low for the shallow interferometers of up to 7 layers we have simulated.

We have implemented the Linear-Optical Feynman Path (LOFP) algorithm with tensor contraction in C, making the code freely available in a public repository. We provide numerical calculations that highlight some cases of interest, where our proposed algorithm performs better against the alternatives. The algorithm is specially useful for shallow interferometers with only a few BS layers, performing better than alternative algorithms for strong simulation, based on Ryser's formula for the calculation of permanents, and also the Cifuentes-Parrilo algorithm which takes into account the interferometer design.

The Feynman path sum technique we implement here can also be applied to simulate certain nonlinear optical gates at essentially no additional cost. For example, nonlinear phase gates are also photon-number preserving, so their simulation involves the same number of Feynman path amplitudes as required for the simulation of linear optics. We leave the investigation of the simulation of more general optical processes for future work.

Other possible future developments building on the work reported here could include code parallelization and simulation capabilities for scenarios involving partially indistinguishable photons, and code implementation of simulation of alternative interferometer designs featuring non-local connections among BS layers. It could also be interesting to consider using Feynman paths to simultaneously calculate and store the amplitudes of all possible output states for a given input, a scenario where our approach might also be advantageous against other alternatives.

\section*{Acknowledgements} 
We would like to thank Raffaele Santagati for helpful discussions at the earlier stages of this project. We acknowledge financial support from H2020-FETOPEN Grant PHOQUSING (GA no.: 899544); from FCT – Fundaç\~{a}o para a Ciência e a Tecnologia (Portugal) via project CEECINST/00062/2018; from the National Council for
Scientific and Technological Development – CNPq (Brazil) under grant 308292/2025-1; and from the Instituto Nacional de Ciência e Tecnologia de Informação Quântica (INCT-IQ).

\bibliographystyle{apsrev}
\bibliography{ernesto_large}


\appendix

\section{Amplitudes for a single beam splitter \label{sec:appendix}}

Here, we present a method for calculating single BS quantum amplitudes for arbitrary Fock state inputs and outputs, which has a runtime that scales linearly with the total number of photons. We will use the fact that the matrix we must take the permanent of contains only 4 distinct elements and has many repeated columns and rows. Using the notation used throughout the paper, the problem is to calculate the elements of the tensor

\begin{equation}
    \bra{y_{1}, y_{2}} BS \ket{x_{1},x_{2}}.
    \label{Eq::1}
\end{equation}
\noindent $x_1$ ($y_1$) and $x_2$ ($y_2$) represent the number of input (output) photons in the up and down waveguides of the BS, respectively.

From \cite{scheel2004permanents}, the value of each element in terms of matrix permanents is given by

\begin{equation}
    \bra{y_{1}, y_{2}} BS \ket{x_{1},x_{2}} = \frac{1}{\sqrt{x_{1}!x_{2}!y_{1}!y_{2}!}}\text{Per}\begin{pmatrix} 
    U_{11}  & \dots & U_{12}  \\
    \vdots & \ddots & \\
    U_{21} &        & U_{22},
    \end{pmatrix}
\end{equation}

\noindent where the elements $U_{ij}$ are the elements of the 2x2 Unitary matrix describing the BS. These are the only unique elements appearing in the matrix, with the input/output states giving the repetitions. The first row will have $U_{11}$ appearing $x_{1}$ times followed by $U_{12}$ $x_{2}$ times. This row is then repeated $y_{1}$ times with the remaining $y_{2}$ rows being made up of  $U_{21}$ appearing $x_{1}$ times and $U_{22}$ $x_{2}$ times. We now appeal to the definition of the permanent for an $N\times N$ matrix $A$

\begin{equation}
    \text{Per}(A) = \sum_{\sigma \in S}\prod_{i=1}^{N} a_{i\sigma(i)}
\end{equation}

\noindent where $S$ is the set of all permutations of  $\{ 1,2,...,N\}$. We can interpret the product term within the sum as being obtained by moving down the rows and selecting columns such that no column is selected more than once and multiplying the corresponding matrix elements together. For a given $\sigma$ if $t$ is the number of times that columns containing $U_{11}$ are selected in the first $y_{1}$ rows, then the product term in the sum will have the form 

\begin{equation}\label{term}
    U_{11}^{t}U_{12}^{(y_{1}-t)}U_{21}^{(x_{1}-t)}U_{22}^{(x_{2} - y_{1} + t)}.
\end{equation}

The parameter $t$ is bounded since we can only select $U_{11}$ columns up to $x_{1}$ or $y_{1}$ times. We can apply the same reasoning to each exponent in (\ref{term}), giving the following restriction on $t$

\begin{align*} 
0 \leq t &\leq  \min(x_{1},y_{1}) \\ 
0 \leq y_{1} - t &\leq  \min(x_{2},y_{1}) \\
0 \leq x_{1} - t &\leq  \min(x_{1},y_{2}) \\
0 \leq x_{2} - y_{1} + t &\leq  \min(x_{2},y_{2}).
\end{align*}

Which, with some algebra, can be arranged to give the following bounds for t

\begin{equation}
    \max (0,y_{1}-x_{2},x_{1}-y_{2}) \leq t \leq \min(x_{1},y_{1}).
\end{equation}

The permanent will then be given by summing \ref{term} over $t$. However, we must also include a combinatorial factor for all the different ways we can obtain the same term. If we select from columns with $U_{11}$ $t$ times and $U_{12}$ columns $y_{1} - t$ times then we can do this in $\frac{x_{1}!}{t!(y_{1}-t)!}$ ways, we also have an additional $y_{1}!$ factor for the choice of columns. Similarly for the $U_{21},U_{22}$ columns we would have $\frac{y_{2}!x_{2}!}{(x_{1}-t)!(y_{1}-x_{2}+t)!}$. This leads us to the following formula for BS permanents, which contains $\mathcal{O}(N)$ terms where each successive term is related to the next by a multiplicative factor. Hence the run time for BS permanents scales as $\mathcal{O}(N)$:

\begin{equation}
    \begin{split}
        \sum_{t} \frac{y_{1}!x_{1}!}{t!(y_{1}-t)!}\frac{y_{2}!x_{2}!}{(x_{1}-t)!(y_{1}-x_{2}+t)!}& \\ U_{11}^{t}U_{12}^{(y_{1}-t)}U_{21}^{(x_{1}-t)}U_{22}^{(x_{2} - y_{1} + t)}.
    \end{split}
\end{equation}

\end{document}